\documentclass[10pt]{article}

\usepackage{times}
\usepackage{amsmath, amssymb}
\usepackage{xspace}
\usepackage{mathpartir}
\usepackage[only, mapsfrom, llbracket, rrbracket]{stmaryrd}
\usepackage{xypic}
\usepackage{listings}

\newcommand{\II}{\mathbb{I}} 
\newcommand{\JJ}{\mathbb{J}} 
\newcommand{\OO}{\mathbb{O}} 
\newcommand{\ZZ}{\mathbb{Z}} 

\newcommand{\Eff}{\textit{Eff}\xspace}
\newcommand{\eff}{\textit{eff}\xspace}

\newcommand{\set}[1]{\{#1\}}

\newcommand{\lift}[1]{#1^\dagger}

\newcommand{\case}{\mathop{\text{\texttt{|}}}}

\newcommand{\bnfis}{\mathrel{\;{:}{:}\!=}\;}
\newcommand{\bnfor}{\mathrel{\;\big|\;}}
\newcommand{\ctx}{\Gamma}
\newcommand{\ent}[1][]{\vdash_{\!#1}}
\newcommand{\ente}{\ent[\mathsf{e}]}
\newcommand{\entc}{\ent[\mathsf{c}]}
\newcommand{\T}{\mathop{:}}

\newcommand{\type}[1]{\mathtt{#1}}
\newcommand{\effect}[1]{\mathtt{#1}}
\newcommand{\emptyty}{\type{empty}}
\newcommand{\unitty}{\type{unit}}
\newcommand{\intty}{\type{int}}
\newcommand{\boolty}{\type{bool}}
\newcommand{\hto}{\Rightarrow}

\newcommand{\op}{\mathtt{op}}
\newcommand{\hash}[2]{#1\,\text{\texttt{\char35}}\,#2}

\newcommand{\kord}[1]{\mathtt{#1}}
\newcommand{\kop}[1]{\;\mathtt{#1}\;}
\newcommand{\kpre}[1]{\mathtt{#1}\;}
\newcommand{\kpost}[1]{\;\mathtt{#1}}
\newcommand{\unt}{\kord{\text{\texttt{()}}}}

\newcommand{\absurd}[1]{\kpre{match} #1 \kpost{with}}
\newcommand{\matchsum}[5]{\kpre{match} #1 \kop{with} \Left{#2} \mapsto #3 \case \Right{#4} \mapsto #5}
\newcommand{\matchpair}[3]{\kpre{match} #1 \kop{with} \pair{#2} \mapsto #3}

\newcommand{\tru}{\kord{true}}
\newcommand{\fls}{\kord{false}}

\newcommand{\Left}[1]{\kpre{Left} #1}
\newcommand{\Right}[1]{\kpre{Right} #1}

\newcommand{\pair}[1]{\text{\texttt{(}}#1\text{\texttt{)}}}

\newcommand{\fun}[1]{\kpre{fun} #1 \mapsto}
\newcommand{\handler}{\kpre{handler}}
\newcommand{\val}{\kpre{val}}
\newcommand{\fin}{\kpre{finally}}

\newcommand{\new}{\kpre{new}}
\newcommand{\newwith}[3]{\kpre{new} #1 \mathrel{@} #2 \kop{with} #3 \kpost{end}}

\newcommand{\ifthenelse}[3]{\kpre{if} #1 \kop{then} #2 \kop{else} #3}
\newcommand{\handle}[2]{\kpre{with} #2 \kop{handle} #1}
\newcommand{\letin}[1]{\kpre{let} #1 \kop{in}}
\newcommand{\letrecin}[1]{\kpre{let} \kpre{rec} #1 \kop{in}}

\newcommand{\lookup}[2]{#1[#2]}
\newcommand{\change}[4]{#1[#2] \mapsfrom #3 \,;\, #4}

\newcommand{\sem}[2]{\llbracket #1 \rrbracket \, {#2}}

\newcommand{\Evalsym}{\mathcal{E}}
\newcommand{\Eval}[1]{\Evalsym(#1)}

\newcommand{\retract}[4]{\xymatrix{*!R{#1\;} \ar@<0.25em>[r]^{#3} & *!L{\;#2} \ar@<0.25em>[l]^{#4}}}

\newcommand{\sto}{\mapsto}
\newcommand{\extend}[3]{#1[#2 \sto #3]}
\newcommand{\extends}[2]{#1[#2]}

\lstnewenvironment{source}{\lstset{
basicstyle=\ttfamily\small,
}}{}
\let\inline\lstinline

\usepackage[
  pdfauthor={Andrej Bauer and Matija Pretnar},
  pdftitle={Programming with Algebraic Effects and Handlers}
]{hyperref}

\begin{document}

\title{Programming with \\ Algebraic Effects and Handlers}

\author{
Andrej Bauer\\\texttt{\small andrej@andrej.com}
\and
Matija Pretnar\\\texttt{\small matija@pretnar.info}
}
\date{\small Department of Mathematics and Physics\\
University of Ljubljana, Slovenia}

\maketitle


\begin{abstract}
  \Eff is a programming language based on the algebraic approach to computational effects,
  in which effects are viewed as algebraic operations and effect handlers as homomorphisms
  from free algebras. \Eff supports first-class effects and handlers through which we may
  easily define new computational effects, seamlessly combine existing ones, and handle
  them in novel ways. We give a denotational semantics of \eff and discuss a prototype
  implementation based on it. Through examples we demonstrate how the standard effects
  are treated in \eff, and how \eff supports programming techniques that use various forms
  of delimited continuations, such as backtracking, breadth-first search, selection
  functionals, cooperative multi-threading, and others.
\end{abstract}

\section*{Introduction}
\label{sec:introduction}

\Eff is a programming language based on the algebraic approach to effects, in
which computational effects are modelled as operations of a suitably chosen
algebraic theory~\cite{plotkin03algebraic}. Common computational effects such as
input, output, state, exceptions, and non-determinism, are of this kind.
Continuations are not algebraic~\cite{hyland07combining}, but they turn out to
be naturally supported by \eff as well. Effect handlers are a related
notion~\cite{plotkin09handlers,pretnar10:_logic_handl_algeb_effec} which
encompasses exception handlers, stream redirection, transactions, backtracking,
and many others. These are modelled as homomorphisms induced by the universal
property of free algebras.

Because an algebraic theory gives rise to a monad~\cite{plotkin02notions},
algebraic effects are subsumed by the monadic approach to computational
effects~\cite{benton00monads}. They have their own virtues, though. Effects are
combined more easily than monads~\cite{hyland06combining}, and the interaction
between effects and handlers offers new ways of programming. An experiment in
the design of a programming language based on the algebraic approach therefore
seems warranted.

Philip Wadler once opined~\cite{wadler95monads} that monads as a programming
concept would not have been discovered without their category-theoretic
counterparts, but once they were, programmers could live in blissful ignorance of
their origin. Because the same holds for algebraic effects and handlers, we
streamlined the paper for the benefit of programmers, trusting that connoisseurs
will recognize the connections with the underlying mathematical theory.

The paper is organized as follows. Section~\ref{sec:syntax} describes the syntax
of \eff, Section~\ref{sec:eff-specific} informally introduces constructs
specific to \eff, Section~\ref{sec:type-checking} is devoted to type checking,
in Section~\ref{sec:semantics} we give a domain-theoretic semantics of~\eff, and
in Section~\ref{sec:implementation} we briefly discuss our prototype
implementation. The examples in Section~\ref{sec:examples} demonstrate how
effects and handlers can be used to produce standard computational effects, such
as exceptions, state, input and output, as well as their variations and
combinations. Further examples show how \eff's delimited control capabilities
are used for nondeterministic and probabilistic choice, backtracking, selection
functionals, and cooperative multithreading. We conclude with thoughts about the
future work.

The implementation of \eff is freely available at \url{http://math.andrej.com/eff/}.

\paragraph{Acknowledgements}
\label{sec:acknowledgment}

We thank Ohad Kammar, Gordon Plotkin, Alex Simpson, and Chris
Stone for helpful discussions and suggestions. Ohad Kammar contributed parts of
the type inference code in our implementation of \eff. The cooperative
multithreading example from Section~\ref{sec:cooperative-multithreading} was
written together with Chris Stone.


\section{Syntax}
\label{sec:syntax}

\Eff is a statically typed language with parametric polymorphism and type
inference. Its types include products, sums, records, and recursive type
definitions. To keep to the point, we focus on a core language with
monomorphic types and type checking. The concrete syntax follows that of OCaml~\cite{OCaml}, and except for new constructs, we discuss it only briefly.

\subsection{Types}
\label{sec:types}

Apart from the standard types, \eff has \emph{effect types $E$} and \emph{handler
  types $A \hto B$}:
\begin{align*}
  \tag{type}
  A, B, C \bnfis {}
    &\intty \bnfor
    \boolty \bnfor
    \unitty \bnfor
    \emptyty \bnfor
    \\
    &A \times B \bnfor
    A + B \bnfor
    A \to B \bnfor
    E \bnfor
    A \hto B,\\
  \tag{effect type}
  E \bnfis {}
    &\kpre{effect} (\kpre{operation} \op_i \T A_i \to B_i)_i \kpost{end}.
\end{align*}
In the rule for effect types and elsewhere below $(\cdots)_i$ indicates that
$\cdots$ may be repeated a finite number of times. We include the empty type
as we need it to describe exceptions, see Section~\ref{sec:exceptions}.
An effect type describes a collection of related operation symbols, for example those for
writing to and reading from a communication channel. We write $\op \T A \to B \in E$ or
just $\op \in E$ to indicate that the effect type $E$ contains an operation $\op$ with parameters of
type $A$ and results of type $B$.
The handler type $A \hto B$ should be understood as the type of handlers acting on
computations of type~$A$ and yielding computations of type~$B$.

\subsection{Expressions and computations}

\Eff distinguishes between \emph{expressions} and \emph{computations}, which are
similar to values and producers of fine-grain call-by-value~\cite{levy03modelling}. The former are inert and free from computational
effects, including divergence, while the latter may diverge or cause
computational effects. As discussed in Section~\ref{sec:implementation}, the
concrete syntax of \eff hides the distinction and allows the programmer to
freely mix expressions and computations.

Beware that we use two kinds of vertical bars below: the tall~$\bnfor$ separates
grammatical alternatives, and the short~$\case$ separates cases in handlers and
match statements. The expressions are
\begin{align*}
  \tag{expression}
  e \bnfis {}
    &x \bnfor
    n \bnfor
    c \bnfor
    \tru \bnfor
    \fls \bnfor
    \unt \bnfor
    \pair{e_1, e_2} \bnfor \\
    &\Left{e} \bnfor \Right{e} \bnfor
    \fun{x \T A} c \bnfor
    \hash{e}{\op} \bnfor
    h, \\
  \tag{handler}
  h \bnfis {}
    &\handler
    (\hash{e_i}{\op_i} \, x \, k \mapsto c_i)_i \case
    \val x \mapsto c_v \case
    \fin x \mapsto c_f,
\end{align*}
where $x$ signifies a variable, $n$ an integer constant, and $c$ other built-in constants.
The expressions $\unt$, $\pair{e_1, e_2}$, $\Left{e}$, $\Right{e}$, and $\fun{x \T A} c$
are introduction forms for the unit, product, sum, and function types, respectively.
Operations $\hash{e}{\op}$ and handlers $h$ are discussed in
Section~\ref{sec:eff-specific}.

The computations are
\begin{align*}
  \tag{computation}
  c \bnfis {}
    &\val e \bnfor
    \letin{x = c_1} c_2 \bnfor
    \letrecin{f \, x = c_1} c_2 \bnfor \\
    &\ifthenelse{e}{c_1}{c_2} \bnfor
    \absurd{e} \bnfor
    \matchpair{e}{x, y}{c} \bnfor \\
    &\matchsum{e}{x}{c_1}{y}{c_2} \bnfor
    e_1 \, e_2 \bnfor \\
    &\new E \bnfor 
    \newwith{E}{e}{
      (\kpre{operation} \op_i \, x \, @ \, y \mapsto c_i)_i
    } \bnfor \\
    &\handle{c}{e}.
\end{align*}
An expression $e$ is promoted to a computation with $\val e$, but in the concrete syntax
$\kord{val}$ is omitted, as there is no distinction between expressions and computations.
The statement $\letin{x = c_1}{c_2}$ binds $x$ in $c_2$, and $\letrecin{f \, x =
  c_1}{c_2}$ defines a recursive function $f$ in $c_2$. The conditional statement and the
variations of $\kord{match}$ are elimination forms for booleans, the empty type, products,
and sums, respectively. Instance creation and the handling construct are discussed in Section~\ref{sec:eff-specific}.

Arithmetical expressions such as $e_1 + e_2$ count as computations because the arithmetical
operators are defined as built-in constants, so that $e_1 + e_2$ is parsed as a double
application. This allows us to uniformly treat all operations, irrespective
of whether they are pure or effectful (division by zero).

\section{Constructs specific to \eff}
\label{sec:eff-specific}

We explain the intuitive meaning of notions that are specific to \eff, namely
instances, operations, handlers, and resources.
We allow ourselves some slack in distinguishing syntax from semantics, which is treated in detail in Section~\ref{sec:semantics}.
It is helpful to think of a terminating computation as evaluating either to a value or an operation applied to a parameter.

\subsection{Instances and operations}
\label{sec:instances-operations}

The computation $\new E$ generates a fresh \emph{effect instance} of effect type~$E$.
For example, $\new \effect{ref}$ generates a new reference, $\new \effect{channel}$ a new communication channel, etc.
The extended form of $\kord{new}$ creates an effect instance with an associated \emph{resource},
which determines the default behaviour of operations and is explained separately in
Section~\ref{sec:resources}.

For each effect instance $e$ of effect type $E$ and an operation symbol $\op \in E$ there
is an \emph{operation} $\hash{e}{\op}$, also known as a \emph{generic
  effect}~\cite{plotkin03algebraic}. By itself, an operation is a value, and hence
effect-free, but an applied operation $\hash{e}{\op}\,e'$ is a computational effect
whose ramifications are determined by enveloping handlers and the resource associated with~$e$.

\subsection{Handlers}
\label{sec:handlers}

A handler
\begin{equation*}
 h =
 \handler
 (\hash{e_i}{\op_i} \, x \, k \mapsto c_i)_i \case \val x \mapsto c_v \case \fin x \mapsto c_f
\end{equation*}
may be applied to a computation $c$ with the handling construct
\begin{equation}
  \label{eq:handling}
   \handle{c}{h},
\end{equation}
which works as follows (we ignore the $\fin$ clause for the moment):
\begin{enumerate}
\item If $c$ evaluates to $\val e$, \eqref{eq:handling} evaluates to $c_v$ with $x$ bound to $e$.
\item If the evaluation of $c$ encounters an operation $\hash{e_i}{\op_i} \, e$,
  \eqref{eq:handling} evaluates to $c_i$ with $x$ bound to $e$ and $k$ bound to the
  continuation of $\hash{e_i}{\op_i} \, e$, i.e., whatever remains to be computed after the
  operation. The continuation is delimited by~\eqref{eq:handling} and is handled by~$h$ as
  well.
\end{enumerate}
The $\kord{finally}$ clause can be thought of as an outer wrapper which performs an
additional transformation, so that \eqref{eq:handling} is equivalent to
\begin{equation*}
  \letin{x = (\handle{c}{h'})}{c_f}  
\end{equation*}
where $h'$ is like $h$ without the $\kord{finally}$ clause. Such a wrapper is useful
because we often perform the same transformation every time a given handler is applied.
For example, the handler for state handles a computation by transforming it to a function
accepting the state, and $\kord{finally}$ applies the function to the initial state, see
Section~\ref{sec:state}.

If the evaluation of $c$ encounters an operation $\hash{e}{\op} \, e'$ that is not listed
in $h$, the control propagates to outer handling constructs, and eventually to the
toplevel, where the behaviour is determined by the resource associated with $e$.

\subsection{Resources}
\label{sec:resources}

The construct
\begin{equation*}
  \newwith{E}{e}{(\kpre{operation} \op_i \, x \, @ \, y \mapsto c_i)_i}
\end{equation*}
creates an instance $n$ with an associated \emph{resource}, inspired by coalgebraic
semantics of computational
effects~\cite{power04from,plotkin08:_tensor_comod_model_operat_seman}. A resource carries
a state and prescribes the default behaviour of the operations $\hash{n}{\op_i}$.
The paradigmatic case of resources is the definition of ML-style references, see Section~\ref{sec:state}.

The initial resource state for $n$ is set to $e$.
When the toplevel evaluation encounters an operation
$\hash{n}{\op_i} \, e'$, it evaluates $c_i$ with $x$ bound to $e'$
and $y$ bound to the current resource state. The result must be a pair of values, the first of
which is passed to the continuation, and the second of which is the new resource state.
If $c_i$ evaluates to an operation rather than a pair of values, a runtime error is
reported, as there is no reasonable way of handling it.

In \eff the interaction with the real world is accomplished through built-in resources.
For example, there is a predefined channel instance $\kord{std}$ with operations
$\hash{\kord{std}}{\kord{read}}$ and $\hash{\kord{std}}{\kord{write}}$ whose associated
resource performs actual interaction with the standard input and the standard output.

\section{Type checking}
\label{sec:type-checking}

Types in \eff are like those of ML~\cite{milner97the-definition} in the sense that they do not capture any
information about computational effects.
There are two typing judgements, $\ctx \ente e \T A$ states that expression $e$ has type
$A$ in context $\ctx$, and $\ctx \entc c \T A$ does so for a computation~$c$.
As usual, a context $\Gamma$ is a list of
distinct variables with associated types. The standard typing rules for expressions are:
\begin{mathpar}
  \infer
    {x \T A \in \ctx}
    {\ctx \ente x \T A}

  \infer
    {}
    {\ctx \ente n \T \intty}

  \infer
    {}
    {\ctx \ente \tru \T \boolty}

  \infer
    {}
    {\ctx \ente \fls \T \boolty}

  \infer
    {}
    {\ctx \ente \unt \T \unitty}

  \infer
    {\Gamma \ente e_1 : A \\
     \Gamma \ente e_2 : B}
    {\ctx \ente \pair{e_1, e_2} \T A \times B}

  \infer
    {\Gamma \ente e : A}
    {\Gamma \ente \Left{e} : A + B}

  \infer
    {\Gamma \ente e : B}
    {\Gamma \ente \Right{e} : A + B}

  \infer
    {\ctx, x \T A \entc c \T B}
    {\ctx \ente \fun{x \T A} c \T A \to B}
\end{mathpar}
We also have to include judgements that assign types to other built-in constants.
An operation receives a function type
\begin{mathpar}
  \infer
  {\ctx \ente e \T E \\
    \op \T A \to B \in E}
  {\ctx \ente \hash{e}{\op} \T A \to B}
\end{mathpar}
while a handler is typed by the somewhat complicated rule
\begin{equation*}
  \infer
  {\infer{\ctx \ente e_i \T E_i \\
    \op_i \T A_i \to B_i \in E_i \\\\
    \ctx, x \T A_i, k \T B_i \to B \entc c_i \T B}{} \\
     \ctx, x \T A \entc c_v \T B \\
     \ctx, x \T B \entc c_f \T C}
   {\ctx \ente (\handler
     (\hash{e_i}{\op_i} \, x \, k \mapsto c_i)_i \mid
     \val x \mapsto c_v \mid
     \fin x \mapsto c_f) \T A \hto C}
\end{equation*}
which states that a handler first handles a computation of type $A$ into
a computation of type $B$ according to the $\kord{val}$ and operation clauses,
after which the $\kord{finally}$ clause transforms it further into a computation of type $C$.

The typing rules for computations are familiar or expected. Promotions of expressions and $\kord{let}$ statements are typed by
\begin{mathpar}
  \infer
    {\ctx \ente e \T A}
    {\ctx \entc \val e \T A}

  \infer
    {\ctx \entc c_1 \T A \\
     \ctx, x \T A \entc c_2 \T B}
    {\ctx \entc \letin{x = c_1} c_2 \T B}

  \infer
    {\ctx, f \T A \to B, x \T A \entc c_1 \T B \\
     \ctx, f \T A \to B \entc c_2 \T C}
    {\ctx \entc \letrecin{f\,x = c_1} c_2 \T C}
\end{mathpar}
and various elimination forms are typed by
\begin{mathpar}
  \infer
    {\ctx \ente e \T \boolty \\
     \ctx \entc c_1 \T A \\
     \ctx \entc c_2 \T A}
    {\ctx \entc \ifthenelse{e}{c_1}{c_2} \T A}

  \infer
    {\ctx \ente e \T \emptyty}
    {\ctx \entc \absurd{e} \T A}
      
  \infer
    {\ctx \ente e \T A \times B \\
     \ctx, x \T A, y \T B \entc c \T C}
    {\ctx \entc \matchpair{e}{x,y}{c} \T C}

  \infer
    {\ctx \ente e \T A + B \\
     \ctx, x \T A \entc c_1 \T C \\
     \ctx, y \T B \entc c_2 \T C}
    {\ctx \entc \matchsum{e}{x}{c_1}{y}{c_2} \T C}

  \infer
    {\ctx \ente e_1 \T A \to B \\
     \ctx \ente e_2 \T A}
    {\ctx \entc e_1 \, e_2 \T B}
\end{mathpar}
The instance creation is typed by the rules
\begin{mathpar}
  \infer
    {}
    {\ctx \entc \new E \T E}

  \infer
    {\ctx \ente e : C \\
     \op_i \T A_i \to B_i \in E \\
     \ctx, x \T A_i, y : C \entc c_i : B_i \times C}
    {\ctx \entc \newwith{E}{e}{(\kpre{operation} \op_i \, x \, @ \, y \mapsto c_i)_i} \T E}
\end{mathpar}
The rule for the simple form is obvious, while the one for the extended form checks that
the initial state $e$ has type $C$ and that, for each operation $\op_i \T A_i \to B_i \in
E$, the corresponding computation $c_i$ evaluates to a pair of type $B_i \times C$.

Finally, the rule for handling expresses the fact that handlers are like functions:
\begin{mathpar}
    \infer
    {\ctx \entc c \T A \\
     \ctx \ente e \T A \hto B}
    {\ctx \entc \handle{c}{e} \T B}
\end{mathpar}


\section{Denotational semantics}
\label{sec:semantics}

Our aim is to describe a denotational semantics which explains how programs in \eff are evaluated.
Since the implemented runtime has no type information, we give Curry-style semantics
in which terms are interpreted without being typed.
See the exposition by John Reynolds~\cite{reynolds00themeaning} on how such semantics can
be related to Church-style semantics in which types and typing judgements receive meanings.

We give interpretations of expressions and computations in domains of \emph{values} $V$
and \emph{results} $R$, respectively. We follow Reynolds by avoiding a particular choice
of $V$ and $R$, and instead require properties of $V$ and $R$ that ensure the semantics
works out. The requirements can be met in a number of ways, for example by solving
suitable domain equations or by taking $V$ and $R$ to be sufficiently large universal
domains.

The domain $V$ has to contain integers, booleans, functions, etc.
In particular, we require that $V$ contains the following retracts, where $\II$ is a 
set of effect instances, and $\oplus$ is coalesced sum:
\begin{align*}
  &\retract{\ZZ_\bot}{V}{\iota_{\intty}}{\rho_{\intty}}
  &
  &\retract{\set{0, 1}_\bot}{V}{\iota_{\boolty}}{\rho_{\boolty}}
  &
  &\retract{\set{\star}_\bot}{V}{\iota_{\unitty}}{\rho_{\unitty}}
  \\
  &\retract{\II_\bot}{V}{\iota_{\kord{effect}}}{\rho_{\kord{effect}}}
  &
  &\retract{V \times V}{V}{\iota_{\times}}{\rho_{\times}}
  &
  &\retract{V \oplus V}{V}{\iota_{+}}{\rho_{+}}
  \\
  &\retract{R^V}{V}{\iota_{\to}}{\rho_{\to}}
  &
  &\retract{R^R}{V}{\iota_{\hto}}{\rho_{\hto}}
\end{align*}
As expressions are terminating, the bottom element of $V$ is never used to denote
divergence, but we do use it to indicate ill-formed values and runtime errors.

\newcommand{\operationDomain}{\II \times \OO \times V \times R^V}

The domain
\begin{equation}
  \label{eq:resultDomain}
  (V + \operationDomain)_\bot
\end{equation}
embodies the idea that a terminating computation is either a value or an operation applied to a parameter and a continuation. There are canonical retractions from~\eqref{eq:resultDomain} onto the two summands,
\begin{equation}
  \label{eq:resultDomain-retraction}
  \xymatrix{
     *!R{V\;} \ar@<0.25em>[rr]^(0.3){\iota_{\kord{val}}}
     & &
     {(V + \operationDomain)_\bot}
     \ar@<0.25em>[ll]^(0.7){\rho_{\kord{val}}}
     \ar@<-0.25em>[r]_(0.7){\rho_{\kord{oper}}}
     &
     *!L{\;(\operationDomain)_\bot}
     \ar@<-0.25em>[l]_(0.3){\iota_\kord{oper}}
  }
\end{equation}
A typical element of~\eqref{eq:resultDomain} is either $\bot$, of the form $\iota_{\kord{val}}(v)$ for a unique $v \in V$, or of the form $\iota_{\kord{oper}}(n, \op, v, \kappa)$ for unique $n \in \II$, $\op \in \OO$, $v \in V$, and $\kappa \in R^V$. We require that $R$ contains~\eqref{eq:resultDomain} as a retract:
\begin{equation}
  \label{eq:resultDomain-in-R}
  \retract{(V + \operationDomain)_\bot}{R}{\iota_{\kord{res}}}{\rho_{\kord{res}}}.
\end{equation}
We may define a strict map from~\eqref{eq:resultDomain} by cases, with one case specifying how to map $\iota_{\kord{val}}(v)$ and the other how to map $\iota_{\kord{oper}}(n, \op, v, \kappa)$. For example, given a map $f : V \to R$, there is a unique strict map $\lift{f} : (V + \operationDomain)_\bot \to R$, called the \emph{lifting} of~$f$, which depends on $f$ continuously and satisfies the recursive equations
\begin{align*}
  \lift{f}(\iota_{\kord{val}}(v)) &= f(v),
  \\
  \lift{f}(\iota_{\kord{oper}}(n,\op,v,\kappa)) &=
  \iota_{\kord{oper}}(n, \op, v, \lift{f} \circ \rho_{\kord{res}} \circ \kappa).
\end{align*}

An \emph{environment $\eta$} is a map from variable names to values. We denote by
$\extend{\eta}{x}{v}$ the environment which assigns $v$ to $x$ and otherwise behaves as
$\eta$. An expression is interpreted as a map from environments to values. The standard
cases are as follows:
\begin{align*}
  \sem{x}{\eta} &= \eta(x)
  \\
  \sem{n}{\eta} &= \iota_\intty(\overline{n})
  \\
  \sem{\fls}{\eta} &= \iota_\boolty(0)
  \\
  \sem{\tru}{\eta} &= \iota_\boolty(1)
  \\
  \sem{\unt}{\eta} &= \iota_\unitty(\star)
  \\
  \sem{\pair{e_1, e_2}}{\eta} &= \iota_{\times}(\sem{e_1}{\eta}, \sem{e_2}{\eta})
  \\
  \sem{\Left e}{\eta} &= \iota_{+}(\iota_0(\sem{e}{\eta}))
  \\
  \sem{\Right e}{\eta} &= \iota_{+}(\iota_1(\sem{e}{\eta}))
  \\
  \sem{\fun{x \T A} c}{\eta} &= \iota_{\to}(\lambda v : V \,.\, \sem{c}{\extend{\eta}{x}{v}})  
\end{align*}
Of course, we need to provide the semantics of other built-in constants, too.
The interpretation of $\hash{e}{\op}$ make sense only when $e$ evaluates to an instance, so we define
\begin{equation*}
  \sem{\hash{e}{\op}}{\eta} = 
  \begin{cases}
    \iota_\to(\lambda v : V \,.\,
       \iota_{\kord{res}}(\iota_{\kord{oper}}(n, \op, v, \iota_\kord{res} \circ \iota_\kord{val}))) &
    \text{if $\rho_{\kord{effect}}(\sem{e}{\eta}) = n \in \II$,}\\
    \iota_\to(\lambda v : V \,.\, \bot) &
    \text{if $\rho_{\kord{effect}}(\sem{e}{\eta}) = \bot$.}
  \end{cases}
\end{equation*}
The interpretation of a handler is
\begin{multline*}
  \sem{
    \handler
    (\hash{e_i}{\op_i} \, x \, k \mapsto c_i)_i \case
    \val x \mapsto c_v \case
    \fin x \mapsto c_f}{\eta} = \\ 
  \iota_\hto(\lift{f} \circ \rho_{\kord{res}} \circ  h \circ \rho_{\kord{res}})
\end{multline*}
where $f : V \to R$ is $f(v) = \sem{c_f}{\extend{\eta}{x}{v}}$ and $h : (V + \operationDomain)_\bot \to R$ is characterized as follows:
if one of the $\rho_{\kord{effect}}(\sem{e_i}{\eta})$ is $\bot$ we set $h = \lambda x
\,.\, \bot$, otherwise $\rho_{\kord{effect}}(\sem{e_i}{\eta}) = n_i \in \II$ for all $i$
and then we take the $h$ defined by cases as
\begin{align*}
  h(\iota_\kord{val}(v)) &= \sem{c_v}{\extend{\eta}{x}{v}}
  \\
  h(\iota_{\kord{oper}}(n_i, \op_i, v, \kappa)) &=
  \sem{c_i}{\extends{\eta}{x \sto v, k \sto \kappa}} \qquad \text{for all $i$,}
  \\
  h(\iota_{\kord{oper}}(n, \op, v, \kappa)) &=
  \begin{aligned}[t]
  &\iota_{\kord{res}}(\iota_{\kord{oper}}(n, \op, v, h \circ \rho_{\kord{res}} \circ \kappa))\\
  &\qquad \text{if $(n, \op) \neq (n_i, \op_i)$ for all $i$.}
  \end{aligned}
\end{align*}

We proceed to the meaning of computations, which are interpreted as maps from
environments to results. Promotion of expressions is interpreted in the obvious way as
\begin{equation*}
  \sem{\val e}{\eta} = \iota_{\kord{res}}(\iota_\kord{val}(\sem{e}{\eta}))
\end{equation*}
The $\kord{let}$ statement corresponds to monadic-style binding:
\begin{equation*}
  \sem{\letin{x = c_1}{c_2}}{\eta} =
  \lift{(\lambda v : V \,.\, \sem{c_2}{\extend{\eta}{x}{v}})}
  (\rho_{\kord{res}}(\sem{c_1}{\eta})),
\end{equation*}
A recursive function definition is interpreted as
\begin{equation*}
  \sem{\letrecin{f \, x = c_1} c_2}{\eta} = \sem{c_2}{\extend{\eta}{f}{\iota_\to(t)}}
\end{equation*}
where $t : V \to R$ is the least fixed point of the map
\begin{equation*}
  t \mapsto (\lambda v : V \,.\, \sem{c_1}{\extends{\eta}{f \sto \iota_\to(t), x \sto v}}).
\end{equation*}
The elimination forms are interpreted in the usual way as:
\begin{align*}
  \sem{\ifthenelse{e}{c_1}{c_2}}{\eta} &=
  \begin{cases}
    \sem{c_1}{\eta} & \text{if $\rho_{\boolty}{\sem{e}{\eta}} = 1$} \\
    \sem{c_2}{\eta} & \text{if $\rho_{\boolty}{\sem{e}{\eta}} = 0$} \\
    \bot & \text{otherwise}
  \end{cases}
  \\
  \sem{\absurd e}{\eta} &= \bot
  \\
  \sem{\matchpair{e}{x,y}{c}}{\eta} &=
  \begin{aligned}[t]
  &\sem{c}{\extends{\eta}{x \sto v_0, y \sto v_1}} \\
  &\qquad \text{where $(v_0, v_1) = \rho_\times (\sem{e}{\eta})$}
  \end{aligned}
  \\
  \sem{\matchsum{e}{x}{c_1}{y}{c_2}}{\eta} &=
  \begin{cases}
    \sem{c_1}{\extend{\eta}{x}{v}} & \text{if $\rho_{+}(\sem{e}{\eta}) = \iota_0(v)$} \\
    \sem{c_2}{\extend{\eta}{y}{v}} & \text{if $\rho_{+}(\sem{e}{\eta}) = \iota_1(v)$} \\
    \bot & \text{otherwise}
  \end{cases}
  \\
  \sem{e_1 \, e_2}{\eta} &= \rho_\to(\sem{e_1}{\eta}) (\sem{e_2}{\eta})
\end{align*}  
For the interpretation of $\kord{new}$ we need a way of generating fresh names so that we
may sensibly interpret
\begin{equation*}
  \sem{\new E}{\eta} = \iota_{\kord{res}}(\iota_\kord{val}(\iota_{\kord{effect}}(n)))
  \quad \text{where $n \in \II$ is fresh.}
\end{equation*}
The implementation simply uses a local counter, but a satisfactory semantic solution needs
a model of names, such as the permutation models of Pitts and
Gabbay~\cite{gabbay01a-new-approach}, with $\II$ then being the set of atoms.

Finally, the handling construct is just an application
\begin{equation*}
  \sem{\handle{c}{e}}{\eta} = \rho_{\hto}(\sem{e}{\eta}) (\sem{c}{\eta}).
\end{equation*}

\subsection{Semantics of resources}
\label{sec:semantics-resources}

To model resources, the denotational semantics has to support the mutable nature of resource state, for example by explicitly threading it through the evaluation.
But we prefer not to burden ourselves and the readers with the ensuing technicalities.
Instead, we assume a mutable store $\sigma$ indexed by effect instances which supports the lookup and update operations.
That is, $\lookup{\sigma}{n}$ gives the current contents at location $n \in \II$, while
$\change{\sigma}{n}{v}{x}$ sets the contents at location $n$ to $v \in V$ and yields $x$.

A resource describes the behaviour of operations, i.e., it is a map $\OO \times V \times V \to V \times V$ which computes a value and the new state from a given operation symbol, parameter, and the current state. Thus an effect instance is not just an element of $\II$ anymore, but an element of $\JJ = \II \times (V \times V)^{\OO \times V \times V}$. Consequently in the semantics we replace $\II$ with $\JJ$ throughout and adapt the semantics of $\kord{new}$ so that it sets the initial resource state and gives an element of $\JJ$:
\begin{multline*}
  \sem{\newwith{E}{e}{(\kpre{operation} \op_i \, x \, @ \, y \mapsto c_i)_i}}{\eta} = \\
    \change{\sigma}{n}{\sem{e}{\eta}}
           {\iota_\kord{res}(\iota_\kord{val}(\iota_\kord{effect}(n,r)))}
    \quad \text{where $n \in \II$ is fresh}
\end{multline*}
where $r : \OO \times V \times V \to V \times V$ is defined by
\begin{equation*}
  r(\op, v, s) =
  \begin{cases}
    \rho_{\times}(\rho_{\kord{val}}(\rho_\kord{res}(\sem{c_i}{\extends{\eta}{x \sto v, y \sto s}}))) &
      \text{if $\op = \op_i$ for some $i$,} \\
    \bot & \text{otherwise.}
  \end{cases}
\end{equation*}
If no resource is provided, we use a trivial one:
\begin{equation*}
  \sem{\new E}{\eta} =
  \iota_\kord{res}(\iota_\kord{val}(\iota_\kord{effect}(n, \bot)))
  \quad \quad \text{where $n \in \II$ is fresh.}
\end{equation*}

Finally, to model evaluation at the toplevel, we define $\Evalsym : (V + \operationDomain)_\bot \to V$ by cases:
\begin{align*}
  \Eval{\iota_{\val}(v)} &= v \\
  \Eval{\iota_{\kord{oper}}((n, r), \op, v, \kappa)} &=
  \begin{aligned}[t]
  &\change{\sigma}{n}{s}{\Eval{\rho_\kord{res}(\kappa \, v')}} \\
  &\qquad \text{where $(v', s) = r(\op, v, \lookup{\sigma}{n})$}.
  \end{aligned}
\end{align*}
The meaning of a computation $c$ at the toplevel in the environment $\eta$ (and an
implicit resource state $\sigma$) is $\Eval{\rho_\kord{res}(\sem{c}{\eta})}$.


\section{Implementation}
\label{sec:implementation}

To experiment with \eff we implemented a prototype interpreter whose main evaluation loop
is essentially the same as the denotational semantics described in
Section~\ref{sec:semantics}. Apart from inessential improvements, such as recursive type
definitions, inclusion of $\kord{for}$ and $\kord{while}$ loops, and pattern matching, the
implemented language differs from the one presented here in two ways that make it usable:
we implemented Hindley-Milner style type inference with parametric
polymorphism~\cite{milner78atheory}, and in the concrete syntax we hid the distinction
between expressions and computations. We briefly discuss each.

There are no surprises in the passage from monomorphic type checking to parametric
polymorphism and type inference. The infamous value
restriction~\cite{wright95simpleimperative} is straightforward because the distinction
between expressions and computations corresponds exactly to the distinction between
nonexpansive and expansive terms. In fact, it may be worth investing in an effect system
that would relax the value restriction on those computations
that can safely be presumed pure. Because $\new E$ is a computation, effect instances are
\emph{not} polymorphic, which is in agreement with ML-style references being
non-polymorphic.

A syntactic division between pure expressions and possibly effectful computations is
annoying because even something as simple as $f \, x \, y$ has to be written as
$\letin{g = f\, x} g \, y$, and having to insert $\kord{val}$ in the right places is no
fun either. Therefore, the concrete syntax allows the programmer to arbitrarily mix expressions
and computations, and a desugaring phase appropriately separates the two.

The desugaring process is fairly simple. It inserts a $\kord{val}$ when an expression
stands where a computation is expected. And if a computation stands where an expression is expected, the computation is hoisted to an enclosing $\kord{let}$
statement. Because several computations may be hoisted from a single expression, the
question arises how to order the corresponding $\kord{let}$ statements. For example, $(f\, x, g\, y)$ can be desugared as
\begin{equation*}
  \begin{aligned}
    &\letin{a = f \, x} \\ &\quad \letin{b = g \, y} (a, b)
  \end{aligned}
  \qquad\text{or}\qquad
  \begin{aligned}
    &\letin{b = g \, y} \\ &\quad \letin{a = f \, x} (a, b)
  \end{aligned}
\end{equation*}
The order of $f \, x$ and $g \, y$ matters when both computations cause computational
effects. The desugaring phase avoids making a decision by using the \emph{simultaneous}
$\kord{let}$ statement
\begin{equation*}
  \kpre{let} a = f \, x \kop{and} b = g \, y \kop{in} (a, b)
\end{equation*}
which leaves open the possibility of various compiler optimizations. The prototype simply
evaluates simultaneous bindings in the order they are given, and a command-line
option enables sequencing warnings about possible unexpected order of effects.
It could be argued that the warnings should actually be errors, but we allow some slack until
we have an effect system that can detects harmless instances of simultaneous binding.

For one-off handlers, \eff provides an inline syntax so that one can write

\begin{center}
\begin{tabular}{ccc}
\begin{source}
handle
  $c$
with
  $\cdots$
\end{source}&
\quad instead of \quad \hbox{}&
\begin{source}
with
  handler
    $\cdots$
handle
  $c$
\end{source}
\end{tabular}
\end{center}
Additionally, the \inline{val} and \inline{finally} clauses may be omitted, in which case
they are assumed to be the identities.


\section{Examples}
\label{sec:examples}

In this section we consider a number of examples that demonstrate the possibilities offered by first-class effects and handlers. Functional programmers will notice similarities with the monadic programming style, and continuations aficionados will recognize their favourite tricks as well. The point we are making though is that even though monads and continuations can be simulated in \eff, it is usually more natural to use effects and handlers directly.

\subsection{Choice}
\label{sec:choice}

We start with an example that is infrequently met in practice but is a favourite of theoreticians, namely \emph{(nondeterministic) choice}. A binary choice operation which picks a boolean value is described by an effect type with a single operation \inline{decide}:
\begin{source}
type choice = effect
  operation decide : unit -> bool
end
\end{source}
Let \inline{c} be an effect instance of type \inline{choice}:
\begin{source}
let c = new choice
\end{source}
The computation
\begin{source}
let x = (if c#decide () then 10 else 20) in
let y = (if c#decide () then 0 else 5) in
  x - y
\end{source}
expresses the fact that \inline{x} receives either the value \inline{10} or \inline{20}, and \inline{y} the value \inline{0} or \inline{5}. If we ran the above computation we would just get a message about an uncaught operation \inline{c#decide}. For the computation to actually do something we need to wrap it in a handler. For example, if we want \inline{c#decide} to always choose \inline{true}, we handle the operation by passing \inline{true} to the continuation \inline{k}:
\begin{source}
handle
  let x = (if c#decide () then 10 else 20) in
  let y = (if c#decide () then 0 else 5) in
    x - y
with
| c#decide () k -> k true
\end{source}
The result of course is \inline{10}. A more interesting handler is one that collects all possible results. Because we are going to use it several times, we define a handler (the operator \inline{@} is list concatenation):
\begin{source}
let choose_all d = handler
  | d#decide () k -> k true @ k false
  | val x -> [x]
\end{source}
Notice that the handler calls the continuation \inline{k} twice, once for each choice, and it concatenates the two lists so obtained. It also transforms a value to a singleton list. When we run
\begin{source}
with choose_all c handle
  let x = (if c#decide () then 10 else 20) in
  let y = (if c#decide () then 0 else 5) in
    x - y
\end{source}
the result is the list \inline{[10;5;20;15]}.
Let us see what happens if we use two instances of \inline{choice} with two handlers:
\begin{source}
let c1 = new choice in
let c2 = new choice in
  with choose_all c1 handle
  with choose_all c2 handle
    let x = (if c1#decide () then 10 else 20) in
    let y = (if c2#decide () then 0 else 5) in
      x - y
\end{source}
Now the answer is \inline{[[10;5];[20;15]]} because the outer handler runs the inner one twice, and the inner one produces a list of two possible results each time. If we switch the order of handlers \emph{and} of operations,
\begin{source}
let c1 = new choice in
let c2 = new choice in
  with choose_all c2 handle
  with choose_all c1 handle
    let y = (if c2#decide () then 0 else 5) in
    let x = (if c1#decide () then 10 else 20) in
      x - y
\end{source}
the answer is \inline{[[10;20];[5;15]]}. For a true understanding of what is going on, the reader should figure out why we get a list of \emph{four} lists, each containing two numbers, if we switch only the order of handlers but not of operations.

\subsection{Exceptions}
\label{sec:exceptions}

An exception is an effect with a single operation \inline{raise} with an empty result type:
\begin{source}
type 'a exception = effect
  operation raise : 'a -> empty
end
\end{source}
The parameter of \inline{raise} carries additional data that can be used by an exception handler. The empty result type indicates that an exception, once raised, never yields the control back to the continuation. Indeed, as there are no expressions of the empty type (but there are of course computations of the empty type), a handler cannot restart the continuation of \inline{raise}, which matches the standard behaviour of exception handlers.

In practice, most exception handlers are one-off and are written using the inline syntax discussed in Section~\ref{sec:implementation}. There are also convenient general exceptions handlers, for example,
\begin{source}
let optionalize e = handler
  | e#raise _ _ -> None
  | val x -> Some x
\end{source}
converts a computation that possibly raises the given exception \inline{e} to one that yields an optional result. We can use it as follows:
\begin{source}
with optionalize e handle
  /@computation@/
\end{source}
In ML-style languages exceptions can be raised anywhere because \inline{raise} is polymorphic, whereas in \eff we cannot use $\hash{e}{\kord{raise}} \; e'$ freely because its type is empty, not polymorphic. This is rectified with the convenience function
\begin{source}
let raise e x = match (e#raise x) with
\end{source}
of type $\alpha \kpost{exception} \to \alpha \to \beta$ which eliminates the empty type, so we may use $\kord{raise} \; e \; e'$ anywhere.

Another difference between ML-style exceptions and those in \eff is that the former are like a single instance of the latter, i.e., if we were to mimic ML exceptions in \eff we would need a (dynamically extensible) datatype of exceptions \inline{exc} and a single instance \inline{e} of type $\kord{exc} \kpost{exception}$. The definition of \inline{raise} would be
\begin{source}
let raise x = match (e#raise x) with
\end{source}
and exception handling would be done as usual.
One consequence of this is that in ML it is possible to catch \emph{all} exceptions at once, whereas in \eff locally created exception instances are unreachable, just as local references are in ML. Which brings us to the next example.

\subsection{State}
\label{sec:state}

In \eff state is represented by a computational effect with operations for looking up and updating a value:
\begin{source}
type 'a ref = effect
  operation lookup: unit -> 'a
  operation update: 'a -> unit
end
\end{source}
We refer to instances of type \inline{ref} as \emph{references}. To get the same behaviour as in ML, we handle them with
\begin{source}
let state r x = handler
  | val y -> (fun s -> y)
  | r#lookup () k -> (fun s -> k s s)
  | r#update s' k -> (fun s -> k () s')
  | finally f -> f x
\end{source}
The handler passes the state around by converting computations to functions that accept the state. For example, lookup takes the state \inline{s} and passes it to the continuation \inline{k}. Because $\kord{k}\,\kord{s}$ is handled too, it is again a function accepting state, so we pass \inline{s} to $\kord{k}\,\kord{s}$ again, which explains why we wrote $\kord{k}\,\kord{s}\,\kord{s}$. Values and updates are handled in a similar fashion. The \inline{finally} clause applies the function so obtained to the initial state~\inline{x}.

The above handler is impractical because for every use of a reference we have to repeat the idiom
\begin{source}
let r = new ref in
  with state r x handle
    /@ computation @/
\end{source}
An even bigger problem is that the reference may propagate outside the scope of its handler where its behaviour is undefined, for instance encapsulated in a $\lambda$-abstraction.
The perfect solution to both problems is to use resources as follows:
\begin{source}
let ref x =
  new ref @ x with
    operation lookup () @ s -> (s, s)
    operation update s' @ _ -> ((), s')
  end
\end{source}
With this definition a reference carries a current state which is initially set to \inline{x}, lookup returns the current state without changing it, while update returns the unit and changes the state. With the definition of the operators
\begin{source}
let (!) r = r#lookup ()
let (:=) r v = r#update v
\end{source}
we get \emph{exactly} the ML syntax and behaviour. Of course, a particular reference may still be handled by a custom handler, for example to fetch its initial value from an external persistent storage and save the final value back into it. 

\subsection{Transactions}

We may handle lookup and update so that the state remains unchanged in case an exception occurs. 
The handler which accomplishes this for a given reference \inline{r} is
\begin{source}
let transaction r = handler
  | r#lookup () k -> (fun s -> k s s)
  | r#update s' k -> (fun s -> k () s')
  | val x -> (fun s -> r := s; x)
  | finally f -> f !r
\end{source}
The handler passes around temporary state \inline{s}, just like the \inline{state} handler in Section~\ref{sec:state}, and only commits it to \inline{r} when the handled computation terminates with a value. Thus the computation
\begin{source}
with transaction r handle
  r := 23;
  raise e (3 * !r);
  r := 34
\end{source}
raises the exception \inline{e} with parameter \inline{69}, but does not change the value of \inline{r}.

\subsection{Deferred computations}

There are many variations on store, of which we mention just one that can be implemented with resources, namely \emph{lazy} or \emph{deferred computations}. Such a computation is given by a thunk, i.e., a function whose domain is $\unitty$. If and when its value is needed, the thunk is \emph{forced} by application to \inline{()}, and the result is stored so that it can be given immediately upon subsequent forcing. This idea is embodied in the effect type
\begin{source}
type 'a lazy = effect
  operation force: unit -> 'a
end 
\end{source}
together with functions for creating and forcing deferred expressions:
\begin{source}
type 'a deferred = Value of 'a | Thunk of (unit -> 'a)

let lazy t =
  new lazy @ (Thunk t) with
    operation force () @ v ->
      (match v with
        | Value v -> (v, Value v)
        | Thunk t -> let v = t () in (v, Value v))
  end

let force d = d#force ()
\end{source}
The function \inline{lazy} takes a thunk \inline{t} and creates a new instance whose initial state is \inline{Thunk t}. The first time the instance is forced, the thunk is evaluated to a value \inline{v}, and the state changes to \inline{Value v}. Thereafter the stored value is returned immediately.

If the thunk triggers an operation, \eff reports a runtime error because it does not allow operations in resources. While this may be seen as an obstacle, it also promotes good programming habits, for one should not defer effects to an unpredictable future time. It would be even better if deferred effectful computations were prevented by a typing discipline, but for that we would need an effect system.

\subsection{Input and output}

A program worth running has to connect with the real-world environment in some way. In
\eff this is done cleanly through built-in effect instances that provide an interface to
the operating system. For input and output \eff has a predefined effect type
\begin{source}
type channel = effect
  operation read : unit -> string
  operation write : string -> unit
end
\end{source}
and a \inline{channel} instance \inline{std} which actually writes to standard output and reads from standard input. Of course, we may handle \inline{std} just like any other instance, for example the handler
\begin{source}
handler std#write _ k -> k ()
\end{source}
erases all output, while
\begin{source}
let accumulate = handler
  | std#write x k -> let (v, xs) = k () in (v, x :: xs)
  | val v -> (v, [])
\end{source}
intercepts output and accumulates it in a list so that
\begin{source}
with accumulate handle
  std#write "hello"; std#write "world"; 3 * 14
\end{source}
prints nothing and evaluates to \inline{(42, ["hello"; "world"])}. Similarly, one could
feed the input from a list with the handler
\begin{source}
let read_from_list lst = handler
  | std#read () k -> (fun (s::lst') -> k s lst')
  | val x -> (fun _ -> x)
  | finally f -> f lst
\end{source}
Both handlers can be quite useful for unit testing of interactive programs.

\subsection{Ambivalent choice and backtracking}
\label{sec:amb-backtrack}

We continue with variations of choice from Section~\ref{sec:choice}. Recall that \emph{ambivalent} choice is an operation which selects among several options in such a way that the overall computation succeeds. We first define the relevant types:
\begin{source}
type 'a result = Failure | Success of 'a

type 'a selection = effect
  operation select : 'a list -> 'a
end
\end{source}
The handler which makes \inline{select} ambivalent is
\begin{source}
let amb s = handler
  | s#select lst k ->
    let rec try = function
      | [] -> Failure
      | x::xs -> (match k x with
                    | Failure -> try xs
                    | Success y -> Success y)
    in
      try lst
\end{source}
Given a list of choices \inline{lst}, the handler passes each one to the continuation \inline{k} in turn until it finds one that succeeds. The net effect is a depth-first search with which we may solve traditional problems, such as the 8 queens problem:
\begin{source}
let no_attack (x,y) (x',y') =
  x <> x' && y <> y' && abs (x - x') <> abs (y - y')

let available x qs =
  filter (fun y -> forall (no_attack (x,y)) qs)
         [1;2;3;4;5;6;7;8]

let s = new selection in
with amb s handle
  let rec place x qs =
    if x = 9 then Success qs else
      let y = s#select (available x qs) in
      place (x+1) ((x,y) :: qs)
  in place 1 []
\end{source}
%
%
The function \inline{filter} computes the sublist of those elements in a list that satisfy the given criterion.
The auxiliary function \inline{available} computes a list of available rows in column \inline{x} if queens \inline{qs} have been placed onto the board so far.
As usual, the program places the queens onto the board by increasing column numbers: given a column \inline{x} and the list \inline{qs} of the queens placed so far, an available row \inline{y} is selected for the next queen.
Because the backtracking logic is contained entirely in the handler, we may easily switch from a depth-first search to a breadth-first search by replacing \emph{only} the \inline{amb} handler with
\begin{source}
let bfs s =
  let q = ref [] in
  handler
    | s#select lst k ->
      (q := !q @ (map (fun x -> (k,x)) lst) ;
       match !q with
        | [] -> Failure
        | (k,x) :: lst -> q := lst ; k x)
\end{source}
The \inline{bfs} handler maintains a stateful queue \inline{q} of choice points \inline{(k,x)} where \inline{k} is a continuation and \inline{x} an argument to be passed to it. The \inline{select} operation enqueues new choice points, dequeues a choice point, and activates it.

The fact that \inline{bfs} seamlessly combines a stateful queue with multiple activations of continuations may lure one into writing an imperative solution to the 8 queens problem such as
\begin{source}
let s = new selection in
with amb s handle
  let qs = ref [] in
  for x = 1 to 8 do
    let y = s#select (available x !qs) in
    qs := (x,y) :: !qs
  done ;
  Success !qs
\end{source}
However, because \inline{qs} is handled with a resource \emph{outside} the scope of \inline{amb} a queen once placed onto the board is never taken off, so the search fails. To make sure that \inline{amb} restores the state when it backtracks, the state has to be handled \emph{inside} its scope:
\begin{source}
let s = new selection in
with amb s handle
  let qs = new ref in
  with state qs [] handle
    for x = 1 to 8 do
      let y = s#select (available x !qs) in
      qs := (x,y) :: !qs
    done ;
    Success !qs
\end{source}
The program finds the same solution as the first version. The moral of the story is that even though effects combine easily, their combinations are not always easily understood.

\subsection{Selection functionals}
\label{sec:select-funct}

The \inline{amb} handler finds an answer if there is one, but provides no information on
the choices it made. If we care about the choices that lead to a particular answer we proceed as follows. First we adapt the \inline{select} operation so that it accepts a \emph{choice point} as well as a list of values to choose from:
\begin{source}
type ('a, 'b) selection = effect
  operation select: 'a * 'b list -> 'b
end
\end{source}
The idea is that we would like to record which value was selected at each choice point. Also, multiple invocations of the same choice point should all lead to the same selection.
The handler which performs such a task is
\begin{source}
let select s v = handler
  | s#select (x,ys) k -> (fun cs ->
    (match assoc x cs with
     | Some y -> k y cs
     | None ->
         let rec try = function
           | [] -> Failure
           | y::ys ->
               (match k y ((x,y)::cs) with
                  | Success lst -> Success lst
                  | Failure -> try ys)
         in try ys))
  | val u -> (fun cs ->
      if u = v then Success cs else Failure)
  | finally f -> f [] ;;
\end{source}
The function \inline{assoc} performs lookup in an associative list.
The handler keeps a list \inline{cs} of choices made so far. It handles \inline{select} by reusing a choice that was previously recorded in \inline{cs}, if there is one, or else by
trying in turn the choices \inline{ys} until one succeeds. A value is handled as success if it is the desired one, and as a failure otherwise.

A simple illustration of the handler is a program which looks for a Pythagorean triple:
\begin{source}
let s = new selection in
with select s true handle
  let a = s#select ("a", [5;6;7;8]) in
  let b = s#select ("b", [9;10;11;12]) in
  let c = s#select ("c", [13;14;15;16]) in
    a*a + b*b = c*c
\end{source}
It evaluates to \inline{Success [("c", 13); ("b", 12); ("a", 5)]}.

Mart{\'\i}n Escard{\'o}'s ``impossible'' selection functional~\cite{escardo10selection} may be implemented with our selection handler. Recall that the selection functional $\epsilon$ accepts a propositional function $p : 2^\mathbb{N} \to 2$ and outputs $x \in 2^\mathbb{N}$ such that $p(x) = 1$ if, and only if, there exists $y \in 2^\mathbb{N}$ such that $p(y) = 1$. Such an $x$ can be found by passing to $p$ an infinite sequence of choice points, each selecting either \inline{false} or \inline{true}, as follows:
\begin{source}
let epsilon p =
  let s = new selection in
  let r = (with select s true handle
             p (fun n -> s#select (n, [false; true])))
  in
    match r with
      | Failure -> (fun _ -> false)
      | Success lst ->
        (fun n -> match assoc n lst with
                  | None -> false | Some b -> b)
\end{source}
The \inline{select} handler either fails, in which case it does not matter what we return, or succeeds by computing a list of choices for which \inline{p} evaluates to \inline{true}. In other words, \inline{r} is a basic open neighbourhood on which \inline{p} evaluates to \inline{true}, and we simply return one particular function in the neighbourhood.

There are several differences between our implementation and Escard{\'o}'s Haskell implementation. First, our implementation is \emph{not} recursive, or to be more precise, it only employs structural recursion and whatever recursion is contained in \inline{p}. Second, we compute a basic neighbourhood on which \inline{p} evaluates to \inline{true} and then pick a witness in it, whereas the Haskell implementation directly computes the witness. Third, and most important, we heavily use the intensional features of \eff to direct the search, i.e., we pass a specially crafted argument to \inline{p} which allows us to discover how \inline{p} uses its argument. The result is a more efficient implementation of \inline{epsilon}, which however is not extensional. A Haskell implementation must necessarily be extensional, because all total functionals in Haskell are.

\subsection{Probabilistic choice}

Probabilistic choice is a form of nondeterminism in which choices are made according to a probability distribution. For example, we might define an operation which picks an element from a list according to the given probabilities:
\begin{source}
type 'a random = effect
  operation pick : ('a * float) list -> 'a
end
\end{source}
The operation \inline{pick} accepts a finite probability distribution, represented as a list of pairs whose second components are nonnegative numbers that add up to~$1$. The handler which computes the expected value of a computation of type \inline{float} is fairly simple
\begin{source}
let expectation r = handler
  | val v -> v
  | r#pick lst k ->
      fold_right (fun (x,p) e -> e +. p *. k x) lst 0.0
\end{source}
Here \inline{fold_right} is the list folding operation, e.g., \inline{fold_right f [a;b;c] x} evaluates as \inline{f a (f b (f c x))}.

Computing the distribution of results of a computation is not much more complicated:
\begin{source}
let combine =
  let scale p xs = map (fun (i, x) -> (i, p *. x)) xs in
  let rec add (i,x) = function
    | [] -> [(i,x)]
    | (j,y)::lst ->
      if i = j then (j,x+.y)::lst else (j,y)::add(i,x) lst
  in
    fold_left
      (fun e (d,p) -> fold_right add (scale p d) e) []

let distribution r = handler
  | val v -> [(v, 1.0)]
  | r#pick lst k ->
      combine (map (fun (x,p) -> (k x, p)) lst)
\end{source}
Here, \inline{combine} is the multiplication for the distribution monad that combines a distribution of distributions into a single distribution. The function \inline{map} is the familiar one, while \inline{fold_left} is the left-handed counterpart of \inline{fold_right}.

As an example, let us consider the distribution of positions in a random walk of length $5$, where we start at the origin, and step to the left, stay put, or step to the right with probabilities $2/10$, $3/10$ and $5/10$, respectively. The distribution is computed by
\begin{source}
let r = new random in
let x = new ref in
with distribution r handle
with state x 0 handle
  for i = 1 to 5 do
    x := !x + r#pick [(-1,0.2); (0,0.3); (1,0.5)]
  done ;
  !x
\end{source}
Just like in the 8 queen example from Section~\ref{sec:amb-backtrack} the handler for state must be enclosed by the distribution handler. We were surprised to see that the ``wrong'' order still works:
\begin{source}
let r = new random in
let x = new ref in
with state x 0 handle
with distribution r handle
  for i = 1 to 5 do
    x := !x + r#pick [(-1,0.2); (0,0.3); (1,0.5)]
  done ;
  !x
\end{source}
How can this be? The answer is hinted at by \eff which issues a warning about arbitrary sequencing of effects in the assignment to \inline{x}. If we write the program with less syntactic sugar, we must decide whether to write the body of the loop as
\begin{source}
let a = !x in
let b = r#pick [(-1,0.2); (0,0.3); (1,0.5)] in
  x := a + b
\end{source}
or as
\begin{source}
let b = r#pick [(-1,0.2); (0,0.3); (1,0.5)] in
let a = !x in
  x := a + b
\end{source}
In the first case \inline{a} holds the value of \inline{x} as it is \emph{before}
probabilistic choice happens, so update correctly reinstates the value of \inline{x}, whereas in the second case it fails to do so. Indeed, we get the wrong answer if we swap the summands in the assignment to \inline{x} and handle state on the outside. On one hand we should not be surprised that the order in which effects happen matters, but on the other it is unsatisfactory that a simple change in the order of addition matters so much. Perhaps the sequencing warnings should be errors after all.

\subsection{Cooperative multithreading}
\label{sec:cooperative-multithreading}

Cooperative multithreading is a model for parallel programming in which several threads run in parallel, but only one at a time. A new thread is created with a \inline{fork} operation, a running thread relinquishes control with a \inline{yield} operation, and a scheduler decides which thread runs next. As is well known, cooperative multithreading can be implemented in languages with first-class continuations.

To get cooperative multithreadding in \eff we first define an effect type with the desired operations:
\begin{source}
type coop = effect
  operation yield : unit -> unit
  operation fork : (unit -> unit) -> unit
end
\end{source}
Next we define a scheduler, in our case one that uses a round-robin strategy, as a handler:
\begin{source}
let round_robin c =
  let threads = ref [] in
  let enqueue t = threads := !threads @ [t] in
  let dequeue () =
    match !threads with
    | [] -> ()
    | t :: ts -> threads := ts ; t ()
  in
  let rec scheduler () = handler
    | c#yield () k -> enqueue k ; dequeue ()
    | c#fork t k ->
        enqueue k ; with scheduler () handle t ()
    | val () -> dequeue ()
  in
    scheduler ()
\end{source}
The handler keeps a queue of inactive threads, represented as thunks. Note that dequeuing automatically activates the dequeued thunk. Yield enqueues the current thread, i.e., the continuation, and activates the first thread in the queue. Fork enqueues the current thread and activates the new one (an alternative would enqueue the new thread and resume the current one). The handler must not just activate the newly forked thread but also wrap itself around it, lest \inline{yield} and \inline{fork} triggered by the new thread go unhandled. Thus the definition of the handler is recursive. The \inline{val} clause makes sure that the threads in the queue get a chance to run when the current thread terminates. 

Nothing prevents us from combining threads with other effects: threads may use common or private state, they may raise exceptions, inside a thread we can have another level of multithreading, etc.

\subsection{Delimited control}
\label{sec:delim-cont}

Our last example shows how to implement standard delimited continuations in \eff. As a result we can transcribe code that uses continuations directly into \eff, although we have found that transcriptions are typically cleaner and easier to understand if we modify them to use operations and handlers directly.

We consider the static variant of delimited control that uses \inline{reset} and \inline{shift}~\cite{danvy92representing}. The first operation delimits the scope of the continuation and the second one applies a function to it, from which it follows that one acts as a handler and the other as an operation. Indeed, the \eff implementation is as follows:
\begin{source}
type ('a, 'b) delimited =
effect
  operation shift : (('a -> 'b) -> 'b) -> 'a
end

let rec reset d = handler
  | d#shift f k -> with reset d handle (f k)
\end{source}
Since \inline{f} itself may call \inline{shift}, the handler wraps itself around \inline{f k}.
The standard useless example of delimited control is
\begin{source}
let d = new delimited in
with reset d handle
  d#shift (fun k -> k (k (k 7))) * 2 + 1
\end{source}
The captured continuation \inline{k} multiplies the result by two and adds one, thus the
result is $2 \times (2 \times (2 \times 7 + 1) + 1) + 1 = 63$. In Scheme the obligatory example of (undelimited) continuations is the ``yin yang puzzle'', whose translation in \eff is
\begin{source}
let y = new delimited in
with reset y handle
  let yin  =
    (fun k -> std#write "@" ; k) (y#shift (fun k -> k k))
  and yang =
    (fun k -> std#write "*" ; k) (y#shift (fun k -> k k))
  in
    yin yang
\end{source}
The self-application \inline{k k} is suspect, and \eff indeed complains that it cannot solve the recursive type equation $\alpha = \alpha \to \beta$. We have not implemented unrestricted recursive types, but we can turn off type checking, after which the puzzle
prints out the same answer as the original one in Scheme.


\section{Discussion}
\label{sec:conclusion}
 
Our purpose was to design a programming language based on the algebraic approach to computational effects and their handlers. We feel that we succeeded and that our experiment holds many promises.

First, we already pointed out several times that \eff would benefit from an effect system that provided a static analysis of computational effects. However, for a useful result we need to find a good balance between expressivity and complexity.

Next, it is worth investigating how to best reason about programs in \eff. Because the language has been inspired by an algebraic point of view, it seems clear that we should look into equational reasoning. The general theory has been investigated in some detail~\cite{plotkin08a-logic, pretnar10:_logic_handl_algeb_effec}, but the addition of effect instances may complicate matters.

Finally, continuations are the canonical example of a non-algebraic computational effect, so
it is a bit surprising that \eff provides a flexible and clean form of delimited control, especially since continuations were not at all on our design agenda. What then can we learn from \eff about control operators in an effectful setting?


\bibliographystyle{plain}



\end{document}